\documentclass[conference]{IEEEtran}
\ifCLASSINFOpdf
   \usepackage[pdftex]{graphicx}
   \graphicspath{{../pdf/}{../jpeg/}}
   \DeclareGraphicsExtensions{.pdf,.jpeg,.png}
\else
\fi
\usepackage{amsmath}
\usepackage{amssymb}             
\usepackage{amsfonts}            
\usepackage{mathrsfs}            

\usepackage{array}

\ifCLASSOPTIONcompsoc
  \usepackage[caption=false,font=normalsize,labelfont=sf,textfont=sf]{subfig}
\else
  \usepackage[caption=false,font=footnotesize]{subfig}
\usepackage{fixltx2e}

\usepackage{stfloats}
\begin{document}
%
\title{KRM-based Dialogue Management}

\author{\IEEEauthorblockN{Qu Wenwu}
\IEEEauthorblockA{Hisense Corp\\
ShanDong, China\\
Email: quwenwu@hisense.com}
\and
\IEEEauthorblockN{Chi Xiaoyu}
\IEEEauthorblockA{ Qingdao Institute of\\
Beihang University\\
ShanDong, China\\
Email: terry.chi@goertek.com}
\and
\IEEEauthorblockN{Zheng Wei}
\IEEEauthorblockA{Institute of Software\\
Chinese Academy of Sciences\\
Beijing, China\\
Email: zhengw@ios.ac.cn}}


%


\maketitle

\begin{abstract}
A KRM-based dialogue management (DM) is proposed using to implement human-computer dialogue system in complex scenarios. KRM-based DM has a well description ability and it can ensure the logic of the dialogue process. Then a complex application scenario in the Internet of Things (IOT) industry and a dialogue system implemented based on the KRM-based DM will be introduced, where the system allows enterprise customers to customize topics and adapts corresponding topics in the interaction process with users. The experimental results show that the system can complete the interactive tasks well, and can effectively solve the problems of topic switching, information inheritance between topics, change of dominance. 
\end{abstract}


%
\IEEEpeerreviewmaketitle

\section{Introduction}
%
%

%
%

Dialogue is a kind of behavior of human brain. The brain receives the grammatical expression of information through vision or hearing, recognizes the semantics of information from accumulated knowledge, decides the semantics that need to be replied, expresses the semantics into natural language grammar, and finally outputs it through voice. Dialogue management model is the core technology in the field of human-computer interaction, and the dialogue management module is the core module of a dialogue system. Dialogue management module can identify the user's intention from the input and context information through natural language processing technologies, and then determine the response that the system will output. 

With the development of human-computer interaction technologies, new dialogue management models have been proposed. \cite{WY2015Reveiw} reviews the current dialogue management model, including Finite State Machine based\cite{MTMF1998Diagram,LC2008Dialogue}, Frame based\cite{GD1996Formbased,YY2005Framebased}, Reinforcement Learning based \cite{LE2000Stochastic,SY2008POMDP}, Example-based\cite{LC2006Asituation,KK2010Example}, Plan-based\cite{RA1999Agenda,XW2000Taskbased}, Bayesian Network based\cite{FL2012FST&DBN}, etc. These dialog management models can work well in lots of task-oriented and chat-oriented application scenarios. However, the dialogue process will face many problems, such as topic switching, information inheritance, and change of dominance, in more complex application scenarios, due to the characteristics of human complexity, randomness and irrationality. It requires the dialogue management model to have a better description ability, a higher level of intelligence, so as to be able to solve these problems flexibly and effectively.

Knowledge-Requirement-Memory (KRM) model\cite{QWW2017Intelligence} is a mathematical model proposed to explain the principle of brain information processing. Through the interaction of knowledge, requirement and memory, the KRM model well explains the information process of human brain's abilities including cognition, thinking, creation, language, feeling, etc. The human dialogue behavior is the comprehensive function of these abilities of brain. In the process of dialogue, the brain experiences the process of identifying information, thinking about the knowledge, requirements and memory contained in information, updating requirements, outputting behavioral instructions, and finally result is the dialogue behavior.

This paper will propose a dialogue management model based on KRM model. It is different from the existing dialog management models that it is a requirement based dialog management model. On the one hand, KRM-based DM is suitable for describe complex conversation scenarios, because KRM model is a model used to process general information. On the other hand, in the process of dialogue, the content of requirement, memory and knowledge is dynamic changed, and the final behavior decision is the result of the comprehensive effect of requirement, memory and knowledge, so it is suitable for solving the problems of topic switching, information inheritance between topics, change of dominance.

The organization of this paper is as follows: Chapter 2 introduces the KRM model and KRM-based DM. Chapter 3 introduces a complex application scenario and analyzes the problems with an example. Chapter 4 introduces a system for the complex scenario. Chapter 5 is the conclusion.

\section{KRM-based DM}
\subsection{KRM model}
In the KRM model, three kinds of disjoint information sets are proposed\cite{QWW2017Intelligence}: knowledge, requirement and memory. Among them, knowledge is used to describe the concept definition and ensure that all information processed in the system is semantically defined. A requirement is a conceptual instance of knowledge, which is used to describe the expected information of the system. A memory is also the conceptual instance of knowledge, which is used to describe the information using to match the requirement. The relationship between the three types of information is shown in Figure \ref{fig_KRM}, which is described as five formulas:

%
\begin{figure}[!t]
\centering
\includegraphics[width=3in]{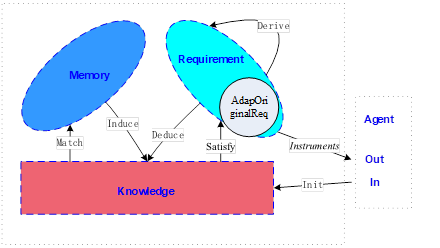}
\caption{Diagram of knowledge, requirement and memory.}
\label{fig_KRM}
\end{figure}


\begin{align}
 &In  = Init(Agent) \\
 &Out = Instruments(Requirement) \\
\begin{split}
\begin{aligned}
&Knowledge  = Generate(In) + Induce(Knowledge,  \\
&\hspace{2em} Memory) + Deduce(Requirement, Knowledge)	
\end{aligned}
\end{split}
\\
\begin{split}
\begin{aligned}
&Requirement = Derive(AdapOriginalReq\&DerivedReq, \\
&\hspace{1.5em}Knowledge)- Satisfy(In, Knowledge, Requirement)
\end{aligned}
\end{split}
\\
&Memory = Match(In, Knowledge, Requirement)
\end{align}

Among them, \textit{In} represents the input information, which is the format information obtained after the initialization (\textit{init}) of the perceptual information. \textit{Out} represents the output information, which is a set of behavior control instructions generated from the \textit{requirement}. \textit{Knowledge} refers to the knowledge set, which can be generated from the input information, induced from the \textit{knowledge} and \textit{memory}, or deduced from the \textit{knowledge} and \textit{requirement}. \textit{Requirement} refers to the requirement set. It has an adaptive original set (\textit{AdapOriginalReq}). A requirement can be derived from existing requirement and knowledge, or be deleted when satisfied. \textit{Memory} refers to the memory set. It is the record of information in the process of information matching.

\subsection{KRM-based DM}
The KRM-based DM includes the management of knowledge, requirement and memory information, as well as the management of information processing.

\subsubsection{Information Neuron}

First, we will introduce a data structure, information neuron. It is an independent information unit integrating information description and calculation.
In the KRM-based DM, information neuron is used to represent knowledge, and an information neuron represents a knowledge, so it is also called knowledge information neuron.
The structure of information neuron is described as follows:

\fbox{\shortstack[l]{
	information neuron = \{\\
	\hspace{2em}ID: the identifier of an information neuron, \\
	\hspace{4em}the ID is globally unique.\\
	\hspace{2em}Name: the name of an information neuron.\\
	\hspace{2em}TriggerInfo: an information used to trigger\\
	 \hspace{4em}the methods in TriggerMethods.\\
	\hspace{2em}Attrs: a set of attributions.\\
	\hspace{2em}Params: a set of parameters.\\
	\hspace{2em}TriggerMethods: a set of methods.\\
	\hspace{2em}StrategyMethods: a set of methods.\}
}}

\textit{ID} and \textit{Name} are assigned when the information neuron is generated. \textit{ID} is the physical index of information neuron, and \textit{Name} is the semantic index of information neuron. When an information neuron is generated, the value of \textit{ID} and \textit{Name} are assigned. There are three kinds of information in information neuron: \textit{TriggerInfo}, \textit{Attrs} and \textit{Params}. \textit{TriggerInfo} describes the triggering condition of an information neuron. \textit{Attrs} is the internal attribute description of knowledge, and \textit{Params} is the parameter description of the attributes to assist in describing the semantics of attributes. There are two kinds of methods in the information neuron. Their trigger condition is different that \textit{TriggerMethods} is triggered by \textit{TriggerInfo}, and \textit{StrategyMethods} is triggered by \textit{Attrs} and \textit{Params}. The implementation of the methods in \textit{TriggerMethods} or \textit{StrategyMethods} can be used to assign values to the attributes and parameters of \textit{Attrs} and \textit{Params}, which can form memory or generate the output to other information neurons.

The structure of requirement and memory is derived from knowledge information neuron, described as follows:

\fbox{\shortstack[l]{
Requirement = \{\\
\hspace{2em}ID\\
 \hspace{2em}Name\\
 \hspace{2em}Reqs = \{attr = valueExpect,param = valueExpect\}\\
\hspace{2em}\}\\
Memory = \{\\
\hspace{2em}ID\\
\hspace{2em}Name\\
\hspace{2em}Mems = \{attr = value, param = value\}\\
\hspace{2em}\}.
}}

\textit{Reqs} represents a set of requirement information and Mems represents a set of memory information. They are consists of a set of \textit{attr} and \textit{param} with values. The difference between them is that the value is a value and the \textit{valueExpect} can be a value or a range of values.

\subsubsection{Information Processing}

The information processing in KRM-based DM is divided into three stages: cognition, understanding and behavior driven. In the cognitive stage, the input grammar information from users is used to trigger corresponding knowledge information neurons. In the understanding stage, the information identified by information neuron is used to generate memory and match requirement. In the behavior driven stage, requirement is used to generate and schedule behavior instructions.

\textbf{Trigger the information neuron}
An index is built for the information neuron, where the information neuron can be retrieved by the \textit{TriggerInfo}. When the \textit{TriggerInfo} of an information neuron is update, the index is also update. In KRM-based DM, the \textit{TriggerInfo} is generate from the input grammar information which ensure that the information neuron is triggered as much as possible.

\textbf{Information identification}
The execution of \textit{TriggerMethods} in an information neurons will assign values to \textit{Attrs} and \textit{Params}. Then the \textit{StrategyMethods} is triggered, and it will also assign values to \textit{Attrs} and \textit{Params}, which is an iterative process inside the information neuron. In addition, the execution of \textit{TriggerMethods} and \textit{StrategyMethods} can output information and trigger other information neurons, which is an iterative process between the information neurons. Because the number of \textit{Attrs} and \textit{Params} is finite, the iteration process will terminate finally. 

\textbf{Match requirement using memory}
When the information content in memory can match information condition of requirement, it is called the requirement is satisfied. Otherwise, the requirement is unsatisfied.

\textbf{Behavior driven by requirement}
Unsatisfied requirements are used to generate behavioral instructions, and then appropriate instructions are scheduled to drive the behavior.

\section{Scenario and Problems}
With the development of internet of things, it is more and more closely combined with artificial intelligence technology. Smart home is one of the most important application scenario. When users have troubles in device installing, using, failure, maintenance, they need to inquire the contact information of the device enterprise and seek the after-sales service. However, with the increasing number and categories of smart devices in the home scenario, the process will bring great inconvenience to the user, because the user may not find the contact information of the enterprise. Therefore, users need a unified entrance to provide services for all devices.

The complex application scenario we will introduce in this chapter is a customer service platform. On the one hand, it allows device enterprises to customize customer service. On the other hand, in the process of interaction with users, it adapts and completes appropriate customer service. For the convenience of description, the service category is limited to device failure reporting service. That is to say, different enterprises can customize the reporting service for different types and different brands, and users will complete any number, any type and any brand of reporting service through an interactive interface.

\begin{table}[!t]
\renewcommand{\arraystretch}{1,3}
\caption{form classification of services}
\label{table_form}
\centering

\begin{tabular}{|c|c|c|c|c|}
\hline
Type & \multicolumn{4}{|c|}{Brand} \\
\hline
Air Conditioner& AC1 & AC2 & AC3 & \\
Refrigerator& RF1 & RF2 & RF3 & RF4\\
Washing Machine& WM1 & WM2 &  & \\
\hline
\end{tabular}
\end{table}

Let's think about how the customer service executive (CSE) will accomplish this task. First of all, the CSE needs to sort out the services by device type and brand, and customize a form of information interaction content for each service (as shown in Table \ref{table_form}).

Then, after the user accesses the call, the CSE will communicate with the user about the type and brand of the device to be reported, and select the corresponding form. According to the content of the form, the CSE will communicate with users about the corresponding information. In this kind of device failure reporting scenario, the device manufacturer needs the user to provide the device information (such as type, brand, failure phenomenon, purchase time) and user information (such as name, phone number, address, appointment time), as well as inform the user of the possible expenses during the maintenance process. In these information, some information must be provided, such as address and appointment time, otherwise the on-site maintenance service cannot be provided. Other information also needs to guide users to provide as much as possible to improve service quality. When the user report multiple device failures, the CSE need to avoid repeated communication of information.

Finally, when all forms are completed, the CSE will end the interaction with the user.

It is a relatively easy task for a person to complete the above task, but there will be many difficulties for an intelligent system. We will analyze the difficulties that need to be overcome in dialogue management through the following specific interaction example.

The difficulties in the above dialogue are listed as follows:

Difficulty 1. The user states the service requirement in U2 that air conditioning has failure. However, there is no device brand information in the information provided by the user, which is not enough to accurately adapt to the customer service customized by the enterprise. How to identify and store the information provided by the user? How to match the customer service information requirements in the subsequent interaction?

Difficulty 2. The user puts forward the failure reporting service of two air conditioners in U4, turns off the existing service in U8, and adds a new service in U10. How to manage the user's reporting service in this complex scenario? How to share or inherit information among services? How to accurately match the information provided by users with the information requirements?

Difficulty 3. The user dominates the topic in U8, U10, etc. How to deal with the dominance?

Difficulty 4. Different interaction strategies will be used for different types of information. How to manage and apply these interaction strategies? E.g. the user's reply to the question in U12 depends on the content of other information, and the user's interaction strategy in U16-20 is based on a finite state machine.

Difficulty 5. When the information provided by the user is added, deleted or updated (U24), how to deal with the information? If there are multiple services, how to update the information of each service?

\fbox{\shortstack[l]{
S1 (S means CSE): Hello, what can I do for you?\\
U2 (U means User): Air conditioner is not cooled.\\
S3: OK. What brand is the air conditioner?\\
U4: there are two air conditioners out of refrigeration. \\
\hspace{2em}One is Hisense and the other is Haier.\\
S5: OK. When did you buy your air conditioners?\\
U6: Hisense was bought the year before last. Haier was\\
\hspace{2em} bought earlier, probably 2011 or 2012.\\
S7: OK. Your air conditioner needs different service \\
\hspace{2em}providers to provide maintenance services. During\\
\hspace{2em} the warranty period, Hisense air conditioner will\\ 
\hspace{2em}charge 30 yuan for on-site maintenance, and no \\
\hspace{2em}additional fees will be charged. When Haier air\\
\hspace{2em} conditioner exceeds the warranty period, it will \\
\hspace{2em}charge 30 yuan for on-site service as well as \\
\hspace{2em}extra maintenance and parts costs.\\
U8: Oh. Haier won't apply for repair, I’ll buy a new one. \\
\hspace{2em}Is the same maintenance service provider only\\
\hspace{2em} charge for one visit?\\
S9: Yes. Please provide your contact number and address.\\
U10: The air deflector of another Hisense air conditioner \\
\hspace{2em}can't move. Can you repair it by the way?\\
S11: Yes. Please provide your contact number and address.\\
U12: When can maintenance personnel come to repair?\\
S13: Please provide your address first, I’ll check the \\
\hspace{2em}schedule of the maintenance personnel in this\\
\hspace{2em} area, and then make an appointment for your repair.\\
U14: My address is No.128 Beijing Road, Qingdao.\\
S15: Is it OK to make an appointment at 3pm tomorrow?\\
U16: I have something to do tomorrow afternoon. \\
\hspace{2em}Tomorrow morning is ok.\\
S17: OK, it's 11am tomorrow. May I have your name? \\
U18: My name is Xie. Could you be earlier?\\
S19: Is it ok at 9am tomorrow?\\
U20: OK.\\
S21: please provide your contact number.\\
U22: 12345678910.\\
S23: OK. Please keep your phone open tomorrow morning.\\
\hspace{2em} The maintenance personnel will contact you by\\
\hspace{2em} phone before coming to the door.\\
U24: OK. If 12345678910 is not answered, you can also\\
 \hspace{2em}call 12345678911.\\
S25: OK. What else can I do for you?\\
U26: No.\\
S27: Have a good life. Bye!
}}	

\section{System Description}
\begin{figure}[!t]
	\centering
	\includegraphics[width=3in]{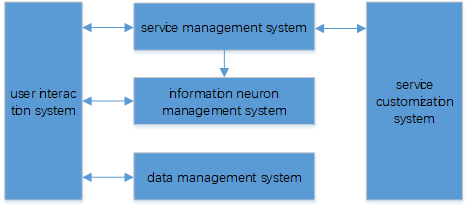}
	\caption{architecture of the system.}
	\label{fig_architecutre}
\end{figure}

The system described in this chapter will complete these two functions: (i) it allows device enterprises to customize failure report service; and (ii) in the process of interaction with users, it adapts and completes appropriate customer service. The system consists of five parts (Fig \ref{fig_architecutre}): service customization system, service management system, information neuron management system, data management system and user interaction system.

Among them, service customization system is responsible for providing service customization function for enterprise. The customized services of enterprise are handled by the service management system, and the cognitive ability of information neuron will be updated with information submitted by enterprise.

The user interaction system is responsible for the information interaction function between the system and the user. It identifies the memory and requirement of the user through the knowledge in the information neuron management system, and then adapts the appropriate service from the service management system, and retrieves the appropriate data from the data management system to satisfy the requirement of the user. The information provided by users in the interaction process is either temporarily stored in the user interaction system, or stored in the data management system for a long time.

\subsection{Information Neuron}			
There are 10 information neuron in the information neuron management system, and their functions are described as follows:

a)	Device information neuron is used to manage device related information. Its attributes include type, brand, failure phenomenon, purchase time, etc.

b)	Device attribute information neuron is used to manage device attribute related information. There are four information neurons in the system: type, brand, failure phenomenon and purchase time.

c)	User information neuron is used to manage user related information. Its attributes include name, address, phone number, appointment time, etc.

d)	User attribute information neuron is used to manage user attribute related information. There are four information neurons: name, address, telephone number and appointment time.

\textbf{Triggering and Identification}
Some of information in \textit{TriggerInfo} is pre-defined manually, and some is supplemented by information submitted by enterprise. Index is built by the information in \textit{TriggerInfo}, through which corresponding information neurons can be triggered.
When an information neuron is triggered, the methods in \textit{TriggerMethods} are executed using to identify the semantic content from grammatical input. In the process of recognition, multiple methods can be used for identifying to assign values to \textit{Attrs} and \textit{Params}. The methods can be based on rules and machine learning.

\begin{figure*}[!t]
	\centering
	\includegraphics[width=6in]{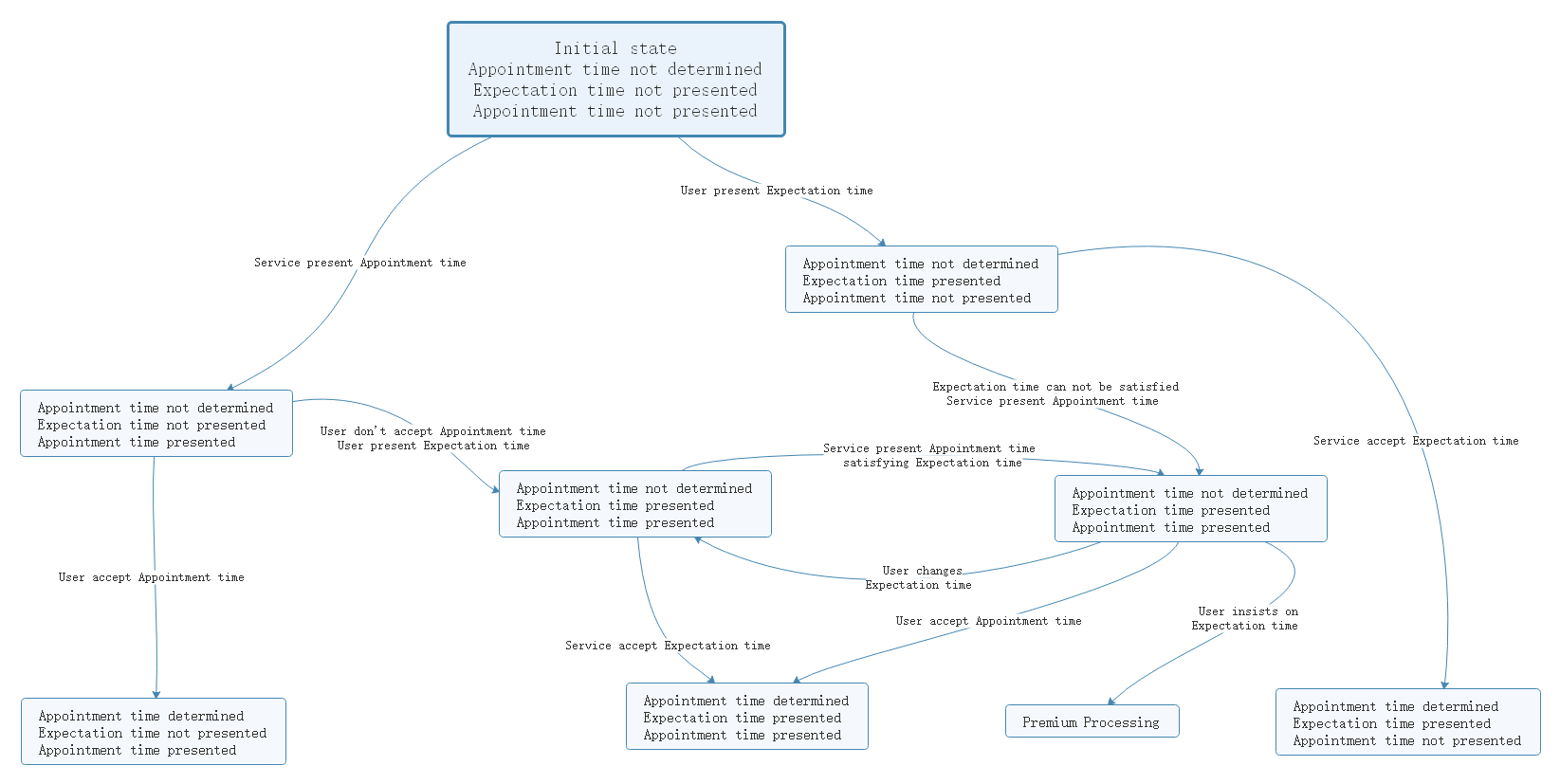}
	\caption{ state transition graph of appointment time.}
	\label{fig_time}
\end{figure*}

\textbf{Strategies of Information Neuron}
When \textit{Attrs} and \textit{Params} in the information neuron are assigned to form memory, the strategies in \textit{StrategyMethods} will be triggered and executed. The strategies are different in different information neuron, and various within the information neuron. E.g.

a)	Update strategy of Attrs and Params.(i) Assign values to phone, address, etc. (ii) When there are multiple devices, update the information memory of corresponding devices through association information and history information.

b)	Generation strategy of candidate reply.
(i) Reply to user's questions. When the user dominates the topic, the user question triggering strategy is used to generate the candidate reply.
(ii) Confirmation of historical information. When the phone number or address have historical information, the confirmation strategy is used to generate the candidate reply.
(iii) Information inquiry. When the system dominates the conversation, the information inquiry strategy is used to generate the candidate reply of phone or address.
(iv) Information interaction. Use the finite state machine (Fig \ref{fig_time}) to generate the candidate reply of appointment time. In the figure, the status is composed of appointment time determined, expectation time presented and appointment time presented. Among them, the appointment time is presented by the server, the expectation time is presented by the user, and the appointment time is determined after being agreed by both (one presented and the other accepted). When the user takes action, the state graph will be migrated accordingly. The system uses the migrated state graph to generate candidate reply.

\subsection{Service Customization and Management}
When enterprise customize the service of device failure reporting, they need to provide the information requirement and the restrictions. The restrictions include the value range and requirement code. E.g. ‘\textit{device brand= Hisense}’ and ‘\textit{requirement code = 1}’ means that the customized service requires the user must provide the device brand information and the device brand must be \textit{Hisense}. 
When the enterprise submits the customized information description, the service management system will generate the corresponding customer service based on the information it provides, and manage the service.

\textbf{Naming Specification of Service}
Each customized service for device failure reporting is called a bot. Each bot has a unique ID in the form of \textit{‘code1\_code2\_code3’}. (i) \textit{code1} is the service type code. E.g. ‘1’ represents device failure reporting service and ‘2’ represents device knowledge question and answer service. (ii) \textit{code2} is the device type code. E.g. ‘1’ represents refrigerator and ‘2’ represents air conditioner.
(iii) \textit{code3} is the device brand code. E.g. ‘1’ represents \textit{Hisense} and ‘2’ represents \textit{Haier}.
This is a tree structure, in which the enterprise customized bots are all leaf nodes, while the middle node is generated by the system to deal with the scenarios where the information provided by users is insufficient to adapt to the customized service. In the system, "0" is used in the bot name to indicate that there is no corresponding code. E.g. ‘1\_1\_0’ represents \textit{‘device failure reporting service\_refrigerator\_NULL’}. In tree structure, if a node has a code that is not 0, it has a parent whose ID is to assign the last non-0 code as 0. E.g. the parent of ‘1\_1\_1’ is ‘1\_1\_0’, and the parent of ‘1\_1\_0’ is ‘1\_0\_0’.

\textbf{Requirements in Bots}
Each bot consists of a \textit{botID}, a set of information requirements and corresponding requirements codes, as shown in the following. The requirement with no value range restriction means any value is satisfied.

\fbox{\shortstack[l]{
Requirement information =  \{\\
	\hspace{2em}botID = 1\_1\_1\\
	\hspace{2em}Equipment brand = Hisense, requirement code = 1\\
	\hspace{2em}Purchase time, requirement code = 0\\
	\hspace{2em}Failure phenomenon, requirement code = 0\\
	\hspace{2em}Name, requirement code = 0\\
	\hspace{2em}Phone, requirement code = 1\\
	\hspace{2em}Address, requirement code = 1\\
	\hspace{2em}Appointment time, requirement code = 1\}	
}}

\textbf{Bot Generation.}
When the enterprise submits the customized service of device failure reporting, the system needs to complete the following processing process to generate a bot. (i) Use the device type and brand information provided by the enterprise to generate botID. (ii) Build and manage the bot. (iii) Use the device information provided by the enterprise to update the \textit{TriggerInfo} of the corresponding information neuron.

Additional, the bot generated by the system initiatively is processed as follows. (i) When the botID generated by the device type and brand information provided by the enterprise does not have a parent node in the tree, the system will generate the parent bot. (ii) The requirement information of the parent bot is the botID information of child nodes. (iii) The bot generated by the system is managed as the same as the bot customized by the enterprise.

\subsection{User Interaction}
\begin{figure*}[!t]
	\centering
	\includegraphics[width=6in]{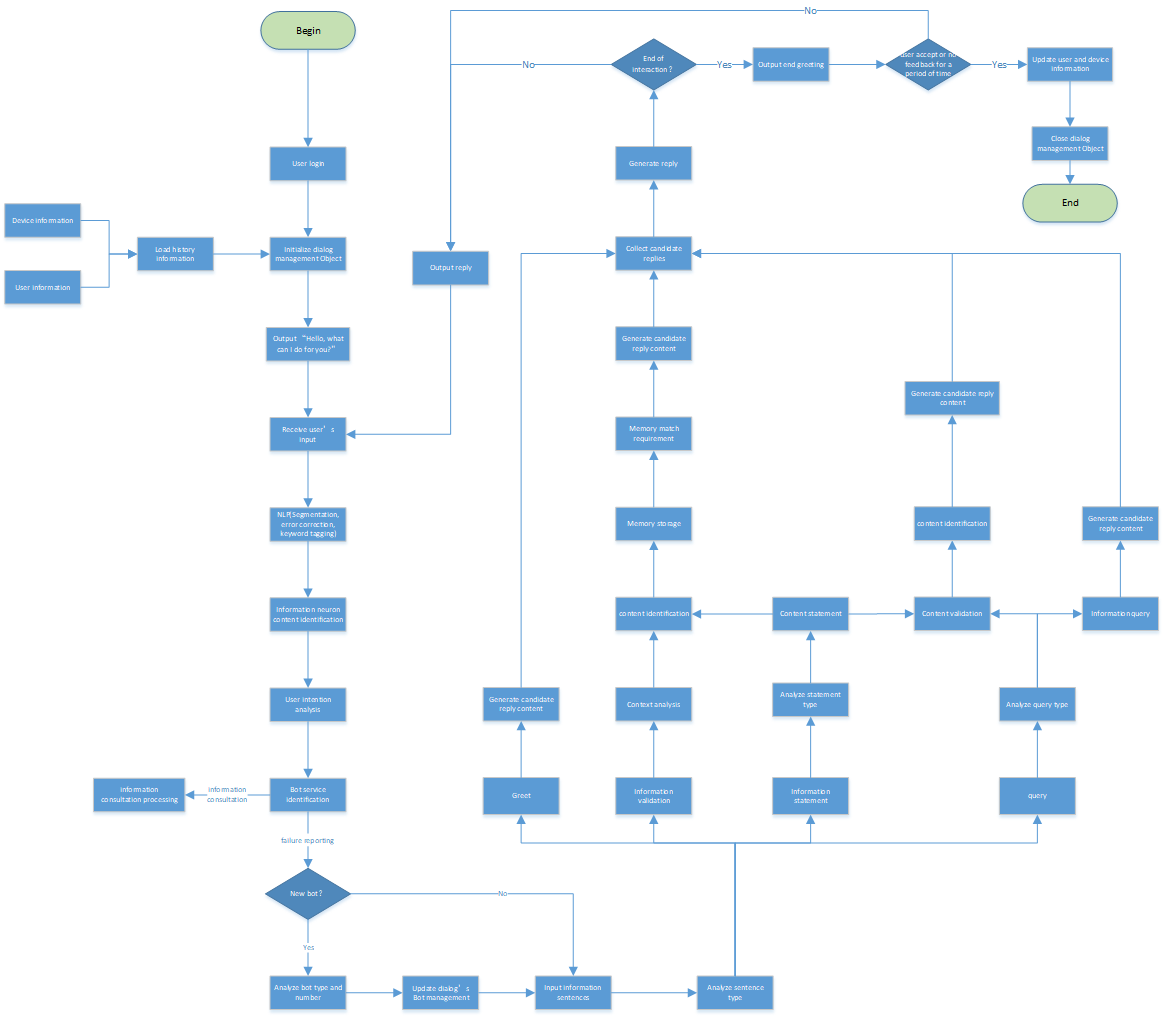}
	\caption{ Information processing flow of user interaction.}
	\label{fig_dialog}
\end{figure*}

The user interaction system is responsible for the natural language interaction with users. When a user logs in to the system, a dialog object will be generated to manage the dialogue with the user. When the dialogue ends, the information obtained in the interaction process will be stored. The information processing flow, shown in Fig \ref{fig_dialog}, includes 8 phases: user login, input preprocessing, semantic information recognition, user requirement analysis, candidate reply generation, memory and requirement matching, reply generation, end of dialogue.

\textbf{User Login.}
When a user logs in, a dialog object will be generated to manage the dialogue with the user. The dialog object first uses the user's identifier to obtain the history information from the data management system, which is used to generate the device information and the user information memory. Then, the system will actively start the interaction with users through greetings.

\textbf{Input Preprocessing.}
The user's input information can be a sentence or a paragraph composed of several sentences. When the system receives the user's input information, it will use natural language preprocessing technology to process the input information, such as error correction and word segmentation.

\textbf{Semantic Information Recognition}
The results of input preprocessing are used to trigger the execution of the methods in the \textit{TriggerMethods} of corresponding information neuron to identify the semantic information. The result of recognition will assign values to the \textit{Attrs} and \textit{Params}, and then trigger the execution of methods in \textit{StrategyMethods}. After finite rounds, this process will end. The \textit{Attrs} and \textit{Params} assigned will be used to generate memory information.

\textbf{User Requirement Analysis.}
The result of input preprocessing and semantic information recognition will trigger the execution of user requirement analysis method to identify whether the user has proposed bot requirements or delete old bot requirements. The dialog object has an active-bot queue that manages the active bots in the user dialog. When a botID is identified, it will be compared with the botID in the active-bot queue. If the botID is a new one, it will be inserted into the active-bot queue. When an old botID is deleted, the botID will be deleted from the active-bot queue.

\textbf{Candidate Reply Generation}
The generation process of candidate replies is described as follows.The first step is to break sentences for user input, because subsequent processing is sentence based. The second step is to analyze the category of each sentence. Each sentence has only one category, and there are four categories of sentences: greetings, information confirmation, information statement and inquiry, according to the priority order of recognition.

a)	Greeting sentence refers to the sentence used for greeting. E.g. ‘Hello’, ‘what can I do for you’.

b)	Information confirmation sentence is the confirmation or denial of the information proposed by the customer service system above. E.g. "OK", "no".

c)	Information statement sentence refers to the sentence used to state information, which is used for user information reply and information coordination between system and users. E.g. ‘12345678910’, ‘ 9am tomorrow’.

d)	Inquiry sentence refers to a sentence with a question word or ending with a question mark. It is used for user information inquiry or information confirmation request. E.g. ‘when can I have on-site service’, ‘3p.m., tomorrow. Is it ok?’

The third step is to process each sentence, using different processing method based on different sentence categories.

a)	Greeting. The system will respond to greetings and guide user topics. The result will be a candidate reply.

b)	Information confirmation. The system will use the information in the context to trigger the corresponding information neuron to complete the semantic information recognition and memory information update. The reply generated by information update triggering method execution will be regarded as the candidate reply.

c)	Information statement. The system will use the sentence and the information in the context to trigger the corresponding information neuron to complete the semantic information recognition and memory information update. The reply generated by information update triggering method execution will be regarded as the candidate reply.

d)	Inquiry. The system identifies the user's requirements through the sentence and context information. For the requirement of information inquiry, the system will retrieve the information to generate a candidate reply. For the requirement of information confirmation, the system will analyze whether the requirement can be satisfied, and the result is used to generate a candidate reply.

\textbf{Memory and Requirement Matching.}
The memory of the device information and the user information are used to match the information requirements in the bot. And then the unsatisfied requirements are used to generate the candidate replies.

\textbf{Reply Generation.}
After the candidate replies are collected, the reply to the user is generated using the established reply strategies. Among them, the priority strategy is described as follows:
a)	The highest priority is bot service confirmation query. If there is a bot service that does not the leaf node, the priority is to guide the user to confirm the leaf bot.
b)	The second is user requirement. If there are user questions, notifications, or unresolved problems, the system needs to satisfy user requirements as early as possible.
c)	The third is to guide the user to provide information of the unsatisfied requirements. (i) First, ask about the fault phenomenon. (ii) Second, confirm the historical information. (iii) Final, the others.

\textbf{End of Dialogue.}
If there is no reply generated from user input, the system will try to end the dialogue. (i)	First, output the closing greeting, such as "what else can I do for you?" (ii) If the user enters a new service requirement or changes the information content of the last service, the input will be processed by the user interaction process. On the contrary, if the user accepts, or there is no response within a period of time, the system ends the dialogue. (iii) Before the end of the dialog, the user information and device failure report information is used to update the data in data management system. And then close the dialog object.

\subsection{System Analysis}
Through the management of knowledge, requirement and memory, the system realizes the dialogue management system for a complex scenario. First of all, we use requirements to describe the topic of interaction scenarios, and construct the relationship between requirements through knowledge information neuron, so as to realize the topic management of complex scenarios. Secondly, the information provided by users can be parsed into memory information through knowledge information neurons, which not only saves the content of information, but also the relationship between information, which lays the foundation for information sharing, information inheritance, memory and requirement matching. Thirdly, through memory information and requirement information, we can decide the content of reply from the overall point of view, so that the result of reply is more reasonable. More detailed information processing of the system is described as follows:

a)	In U2, the user provides \textit{‘device\_type = air conditioner’} and \textit{‘failure\_phenomenon = no cooling’}. Through these two information, the system can infer that the user requires air conditioner failure reporting service. Because there is no brand information, the botID of the leaf node cannot be matched, so the system decides to ask user for the brand of air conditioner. However, the device information provided by the user will be saved in the device information memory.

\fbox{\shortstack[l]{
active-bot queue: "air conditioning failure report"\\
device information 1: type = air conditioning, \\
	\hspace{2em}fault phenomenon = no cooling.
}}

b)	The information stated by the user in U4 will drive the system to adapt to two botID and update to the active-bot queue. And the device information provided by the user will also be saved in two device information memories.

\fbox{\shortstack[l]{
active-bot queue: "Hisense air conditioner reports failure", \\
	\hspace{2em}"Haier air conditioner reports failure"\\
device information 1: type = air conditioner, \\
	\hspace{2em}brand = Hisense, fault phenomenon = no cooling.\\
device information 2: type = air conditioner, \\
	\hspace{2em}brand = Haier, fault phenomenon = no cooling.
}}

c)	The information stated by the user in U6 will be used to update two device information memories.

\fbox{\shortstack[l]{
active-bot queue: "Hisense air conditioner reports failure", \\
	\hspace{2em}"Haier air conditioner reports failure"\\
device information 1: type = air conditioner,\\
	 \hspace{2em}brand = Hisense, fault phenomenon = no cooling, \\
	 \hspace{2em}purchase time = 2017.\\
device information 2: type = air conditioner,\\
	\hspace{2em} brand = Haier, fault phenomenon = no cooling,\\
	\hspace{2em} purchase time = 2011 or 2012.
}}

d)	The information provided by the user in U8 will drive the system to delete "Haier air conditioning failure report" in the active-bot queue. In addition, the dialogue dominance is transferred to the user, so the system's reply content also needs to answer the user's questions first, and then try to dominate the dialogue. In U10, users ignore the topic dominance of the system and continue to ask questions of interest. Here, users add a new air conditioning device failure reporting requirement, and the system also needs to update the active-bot queue and device information memory.

\fbox{\shortstack[l]{
active-bot queue: "Hisense air conditioner reports failure", \\
	\hspace{2em}"Hisense air conditioner reports failure"\\
device information 1: type = air conditioner, \\
	\hspace{2em}brand = Hisense, fault phenomenon = no cooling,\\
	 \hspace{2em}purchase time = 2017.\\
device information 2: type = air conditioner, \\
	\hspace{2em}brand = Haier, fault phenomenon = no cooling,\\
	\hspace{2em} purchase time = 2011 or 2012.\\
device information 3: type = air conditioner,\\
	 \hspace{2em}brand = Hisense, fault phenomenon = air\\
	  \hspace{2em}deflector failure.
}}

e)	The user's dominant topic in U12 is on-site time. The answer to this question depends on the address information, and the system will decide the reply content accordingly. The content stated by the user in u14-u22 will be used to update the user information memory. After the information interaction, the system will guide the user to end the conversation.

\fbox{\shortstack[l]{
user information: address = No.128, Beijing Road,\\
  \hspace{2em}Qingdao City, appointment time = 9:00 a.m. on \\
   \hspace{2em}October 15 2019, name = Xie, Tel. = \\
   \hspace{2em} "12345678910" or "12345678911" (standby).
}}

\section{Conclusion} 
This paper proposes a new KRM-based dialogue management, which can be used to solve more complex interaction scenarios than the traditional ones.

First, the decision-making model based on knowledge, requirement and memory has the better description ability.

Second, information neuron is an independent unit of information storage and calculation. The system model based on information neuron can realize very complex scenarios. Moreover, the independence of information neurons makes it easy to expand the system architecture through distributed methods.

Third, separate memory from requirement. The information provided by users will be transformed into memory. And then memory information will be used to match the requirements of scenarios information, so it is easy to solve the problem of information inheritance in different scenarios.

Fourth, the decision-making process is divided into requirement and memory analysis, memory matching requirement and response decision-making stages, so that the system response no longer only depends on local information, but integrates all information results. Thus, it is easy to solve the problems of user topic switching and dominant change.

%
%
%
%


%
%



%
%
%
%
%
%



\end{document}